\documentclass{aa} 
\usepackage{graphicx}
\usepackage{natbib}
\usepackage{txfonts}

\usepackage[colorlinks]{hyperref}

\hypersetup{linkcolor={blue},citecolor={blue}, urlcolor={blue}
}

\hypersetup{linkcolor={blue},citecolor={blue}, urlcolor={blue}
}

\makeatletter
\renewcommand*\aa@pageof{, page \thepage{} of \pageref*{LastPage}} 
\makeatother

\begin{document} 

   \title{Analysis of the public HARPS/ESO spectroscopic archive}
   \subtitle{Ca~{\sc ii}~H\&K time series for the HARPS radial velocity database\thanks{Tables 1 and 4 are only available in electronic form at the CDS via anonymous ftp to cdsarc.u-strasbg.fr (130.79.128.5) or via http://cdsweb.u-strasbg.fr/cgi-bin/qcat?J/A+A/.}}

   \author{V. Perdelwitz\inst{1,2,5}
          \and
          T. Trifonov\inst{3,4,6}
          \and
          J. T. Teklu\inst{5}
          \and
          K. R. Sreenivas\inst{5}
          \and
          L. Tal-Or\inst{5,7}
          }

   \institute{Department of Earth and Planetary Science, Weizmann Institute of Science, Rehovot, Israel\\
              \email{volker.perdelwitz@weizmann.ac.il}
         \and
             Hamburger Sternwarte, Universität Hamburg, Gojenbergsweg 112, 21029 Hamburg, Germany
         \and
             Department of Astronomy, Faculty of Physics, Sofia University ``St. Kliment Ohridski'', 5 James Bourchier Blvd., BG-1164 Sofia, Bulgaria
        \and
             Max-Planck-Institut für Astronomie, Königstuhl 17, 69117 Heidelberg, Germany             
         \and
             Department of Physics, Ariel University, Ariel 40700, Israel
          \and
             Zentrum f\"ur Astronomie der Universt\"at Heidelberg, Landessternwarte,
              K\"onigstuhl 12, 69117 Heidelberg, Germany
         \and
             Astrophysics Geophysics And Space Science Research Center, Ariel University, Ariel 40700, Israel
             }

   \date{Received 13/10/2023; accepted 16/1/2024}

   \abstract
   {Magnetic activity is currently the primary limiting factor in radial velocity (RV) exoplanet searches. Even inactive stars, such as the Sun, exhibit RV jitter of the order of a few m\,s$^{-1}$ due to active regions on their surfaces. Time series of chromospheric activity indicators, such as the Ca~{\sc ii}~H\&K lines, can be utilized to reduce the impact of such activity phenomena on exoplanet search programmes. In addition, the identification and correction of instrumental effects can improve the precision of RV exoplanet surveys.}
   {We aim to update the {\sc HARPS-RVBank} RV database and include an additional $3.5$ years of time series and Ca~{\sc ii}~H\&K lines ($R_{\mathrm{HK}}^\prime$) chromospheric activity indicators. This additional data will aid in the analysis of the impact of stellar magnetic activity on the RV time series obtained with the HARPS instrument. Our updated database aims to provide a valuable resource for the exoplanet community in understanding and mitigating the effects of such stellar magnetic activity on RV measurements.}
   {The new {\sc HARPS-RVBank} database includes all stellar spectra obtained with the HARPS instrument prior to January 2022. The RVs corrected for small but significant nightly zero-point variations were calculated using an established method. The $R_{\mathrm{HK}}^\prime$ estimates were determined from both individual spectra and co-added template spectra with the use of model atmospheres. As input for our derivation of $R_{\mathrm{HK}}^\prime$, we derived stellar parameters from co-added, high signal-to-noise ratio (S/N) templates for a total of 3\,230 stars using the stellar parameter code {\sc SPECIES}.}
   {The new version of the HARPS RV database has a total of 252\,615 RVs of 5\,239 stars. Of these, 195\,387 have $R_{\mathrm{HK}}^\prime$ values, which corresponds to 77\% of all publicly available HARPS spectra. Currently, this is the largest public database of high-precision (down to $\sim1$\,m\,s$^{-1}$) RVs, and the largest compilation of $R_{\mathrm{HK}}^\prime$ measurements. We also derived lower limits for the RV jitter of F-, G-, and K-type stars as a function of $R_{\mathrm{HK}}^\prime$.}{}

\maketitle

\setlength{\belowcaptionskip}{-5pt}

\section{Introduction}
While the transit method has led to the detection of a greater number of exoplanets compared to the radial velocity (RV) method, the RV technique remains a valuable tool in the detection and characterization of exoplanetary companions. This is because it does not require continuous observations over multiple planetary periods, enabling the discovery of planets with orbital periods of the order of the time span over which spectra have been collected. This time span can be several decades for stars observed by multiple facilities. As such, the RV method enables the creation of a representative sample of planetary companions without the need for continuous coverage.

The yield of a search for substellar companions via the radial velocity (RV) method using high-resolution spectrographs such as the High Accuracy Radial velocity Planet Searcher (HARPS) \citep{2003Msngr.114...20M} is increased by optimizing the RV extraction method \citep[e.g.][]{2018A&A...609A..12Z} and by correcting the RVs for nightly instrumental effects \citep[e.g.][]{2015A&A...581A..38C, 2019MNRAS.484L...8T}. Based on the publicly available HARPS spectra, \cite{2020A&A...636A..74T} published the {\sc HARPS-RVBank}, a catalogue of $>212\,000$ corrected RVs, which have since been used by the community to confirm or discover planet candidates \citep[e.g.][]{2021ApJ...909L...6P,2021arXiv211209029S} and estimate planet occurrence rates \citep{2020A&A...643A.106B}.\\
Aside from the detrimental influence of instrumental effects, stellar magnetic activity has long been known to be the primary cause of RV jitter for a long time \citep{1998ApJ...498L.153S}. In combination with stellar rotation, the presence of active regions on the stellar surface can result in quasi-periodic RV signals mimicking those caused by a substellar companion \citep{2001A&A...379..279Q,2011A&A...525A.140D,2013A&A...552A.103R}. While this RV noise is louder in younger and more active stars \citep{2018A&A...614A.122T, 2019A&A...632A..37B}, it has a more complex dependence on the spectral type \citep{2011A&A...525A.140D}. Even slow rotators with low levels of magnetic activity, such as our Sun, exhibit RV jitter of a few m\,s$^{-1}$ \citep{2015ApJ...798...63M, 2016MNRAS.457.3637H, 2019MNRAS.487.1082C}, thus exceeding the nominal precision of state-of-the-art planet-hunting facilities such as the Echelle Sectrograph for Rocky Exoplanets and Stable Spectroscopic Observations (ESPRESSO) \citep{2014AN....335....8P}, and upcoming facilities.

Though efforts have been made to mitigate the effect of magnetic activity with various methods \citep{2021MNRAS.505.1699C, 2021arXiv211102383B}, the most reliable way to exclude false positive detections to date is to measure the stellar rotation period and demonstrate it to be different from the planetary candidate signal. This can be achieved by analysing photospheric activity indicators \citep[i.e. light curves,][]{2014ApJS..211...24M}, or chromospheric lines such as H$\alpha$ or Ca~{\sc ii}~H\&K \citep[e.g.][]{2021A&A...650A.188A}.
In this publication, we present an update to  the {\sc HARPS-RVBank}, which incorporates newly available observational data, including 252\,615 RVs of 5\,239 stars and measurements of the relative emission in the Ca~{\sc ii}~H\&K lines, $R_{\mathrm{HK}}^\prime$, for 77\% of all spectra.
In Sect.~\ref{sec:2}, we provide details of our re-processing scheme with the Spectrum radial velocity analyser (SERVAL) pipeline, the nightly zero point (NZP) correction process, and the derivation of stellar parameters and R'HK from the public spectra.  In Sect.~\ref{sec:3}, we present our results and provide a discussion on our findings. Finally, in Sect.~\ref{sec:4}, we present a summary and our conclusions.

\section{Updating the {\sc HARPS-RVBank}}
\label{sec:2}
\begin{figure}
  \resizebox{\hsize}{!}{\includegraphics{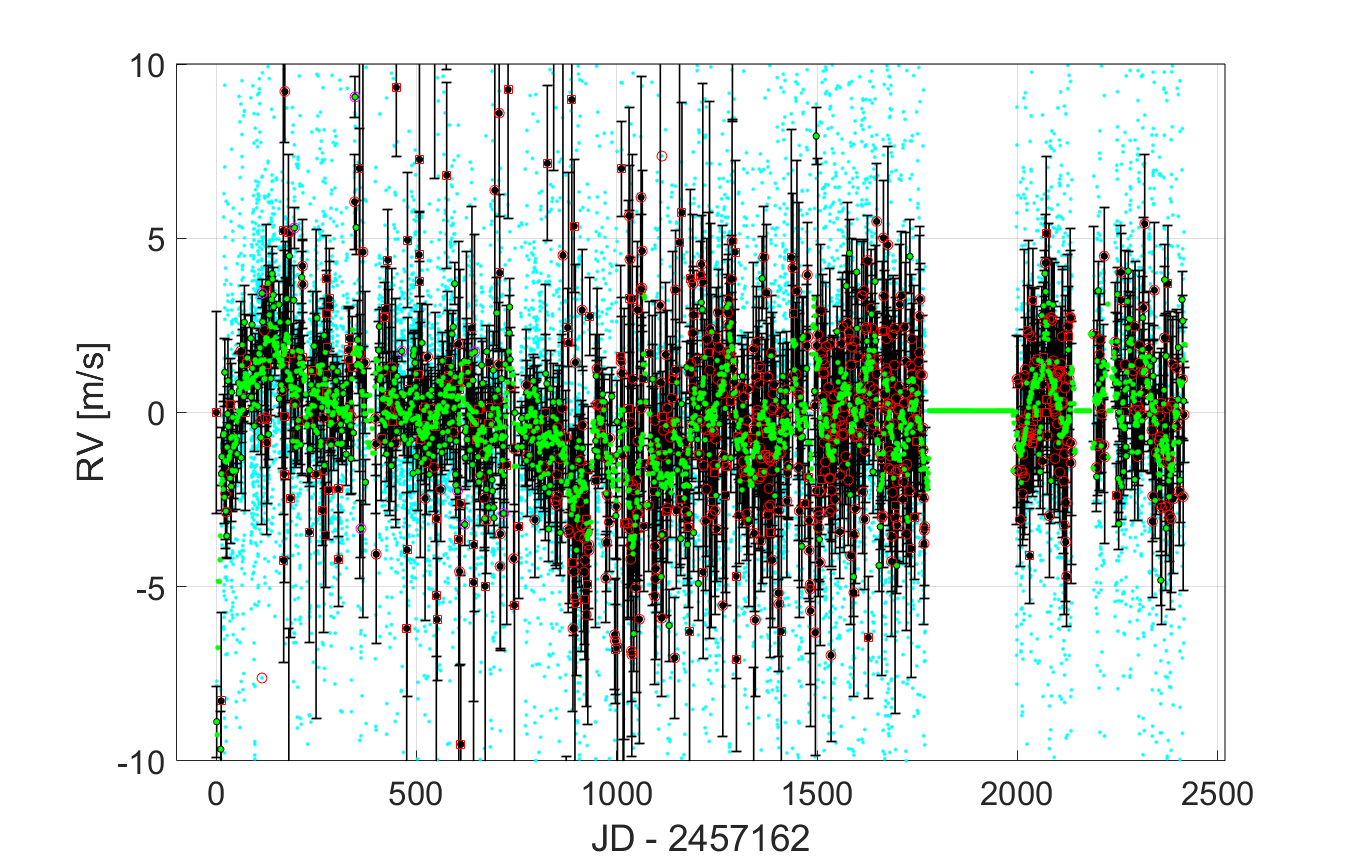}}
  \caption{Temporal evolution of NZPs. \textit{Black errorbars}: HARPS-post SERVAL NZPs. \textit{Cyan dots}: Stellar zero-point subtracted RVs of all RV-quiet stars (weighted RV rms scatter $<10$\,m\,s$^{-1}$). \textit{Red boxes}: NZPs that were calculated with too few ($< 3$) RVs. \textit{Red circles}: NZPs with too large uncertainties ($>1.2$\,m\,s$^{-1}$). \textit{Magenta circles}: Significantly deviating NZPs. The RVs in the red-marked nights were corrected by using a smoothed version of the NZP curve, which was calculated with a moving weighted average (21-d window). For the ten significantly deviating NZPs, we adopted their individual (unsmoothed) NZP values, regardless of their uncertainties. \textit{Green dots}: NZP values that were eventually used for the correction.}
  \label{fig:post-nzp}
\end{figure}

We obtained all available HARPS stellar spectra from the ESO archive (\url{http://archive.eso.org/wdb/wdb/adp/phase3_main/form}) that were observed prior to January 2022. These spectra were processed in the same manner as described in \cite{2020A&A...636A..74T}. In addition to the data reduction software (DRS) products provided by HARPS, we calculated SERVAL RVs and various activity indices using the methodology described in \cite{2018A&A...609A..12Z,2020ascl.soft06011Z}. We also determined the data-driven NZP corrections for both the DRS and SERVAL RVs. 
Figure \ref{fig:post-nzp} shows the zero-point variations of post-SERVAL RVs. The $774$ good NZPs (out of $1395$ nights with useful data) have a weighted rms scatter of $1.8$\,m\,s$^{-1}$ and a median error of $1.1$\,m\,s$^{-1}$. 
Despite the $2020$--$2021$ COVID-related observing gaps, the instrument's zero-point behaviour does not differ much from what was reported in \citet{2020A&A...636A..74T}.

In the first release of the {\sc HARPS-RVBank}, we only included targets with at least three usable spectra, as this was the minimum required for constructing a spectral template and obtaining precise RVs with SERVAL. In this second release of the {\sc HARPS-RVBank}, we extended the inclusion criteria to include targets with a single usable spectrum, significantly increasing the number of targets. While we could not compute meaningful SERVAL RVs and activity indices for these targets, we provide the original HARPS-DRS estimates and used their spectra to estimate Ca~{\sc ii}~H\&K values, which is the primary focus of this work. From 2003 to the present, we retrieved 309,425 unique spectra from the ESO archive. Of those spectra, 12,136 were excluded due to their classification as solar spectra, Solar System objects, quasars (QSOs), and various transients. Additionally, 36,765 spectra were not included in the {\sc HARPS-RVBank} due to failing the quality control check, which includes criteria -- for example -- being taken with an I2 cell, having an extremely low or high signal-to-noise ratio (S/N), or exhibiting other issues that prevent the extraction of meaningful results. The new {\sc HARPS-RVBank} consists of 252\,615 unique spectra of 5\,239 targets\footnote{Instant access to the {\sc HARPS-RVBank} products is made available through the {\sc Exo-Striker} exoplanet toolbox \citep{Trifonov2019_es}, which can be freely accessed at \url{https://github.com/3fon3fonov/exostriker}.} that have passed the semi-automated quality control process \citep[see,][for details]{2020A&A...636A..74T}.

\subsection{Derivation of stellar parameters}
\label{sec:param}

\setlength{\tabcolsep}{3pt}
\renewcommand{\arraystretch}{1.2}
\begin{table*}[ht!]
\centering
\small
\caption{Stellar parameters. The full table is available on the CDS.} \label{table:1}
\begin{tabular}{l c c c c c c c c c c c c c}
\hline
  \multicolumn{1}{c}{Name} &
  \multicolumn{1}{c}{ra} &
  \multicolumn{1}{c}{dec} &
  \multicolumn{1}{c}{T$_{eff}$} &
  \multicolumn{1}{c}{$\Delta$T$_{eff}$} &
  \multicolumn{1}{c}{log\,g} &
  \multicolumn{1}{c}{$\Delta$log\,g} &
  \multicolumn{1}{c}{[Fe/H]} &
  \multicolumn{1}{c}{$\Delta$[Fe/H]} &
  \multicolumn{1}{c}{v\,sini} &
  \multicolumn{1}{c}{$\Delta$v\,sini} &
  \multicolumn{1}{c}{$R_{\mathrm{HK}}^\prime$} &
  \multicolumn{1}{c}{$\Delta R_{\mathrm{HK}}^\prime$} &
  \multicolumn{1}{c}{vflag} \\
& [deg] & [deg]  & [K] & [K] &  &  & [dex] & [dex] & [km\,s$^{-1}$] & [km\,s$^{-1}$] &  &  & \\
\hline
  BD+012494 & 163.0321195 & 0.49329 & 5524 & 50 & 4.1 & 0.1 & 0.29 & 0.042 & 4.0 & 0.3 & 1.07$\cdot 10^{-5}$ & 5.63$\cdot 10^{-7}$ & 0\\
  BD+01316 & 26.63371 & 2.69887 & 6370 & 50 & 4.2 & 0.1 & 0.26 & 0.047 & 5.5 & 0.4 & 6.98$\cdot 10^{-6}$ & 3.59$\cdot 10^{-7}$ & 0\\
  BD+034525 & 318.47394 & 4.45293 & 5754 & 61 & 4.1 & 0.1 & 0.05 & 0.05 & 3.9 & 0.2 & 8.80$\cdot 10^{-5}$ & 1.61$\cdot 10^{-5}$ & 0\\
  BD+053805 & 278.947839 & 5.33738 & 5163 & 50 & 3.0 & 0.1 & -0.09 & 0.043 & 4.2 & 0.4 & 6.49$\cdot 10^{-6}$ & 6.87$\cdot 10^{-7}$ & 0\\
\hline\end{tabular}
\end{table*}

\begin{figure}
  \centering
  \resizebox{\hsize}{!}{\includegraphics{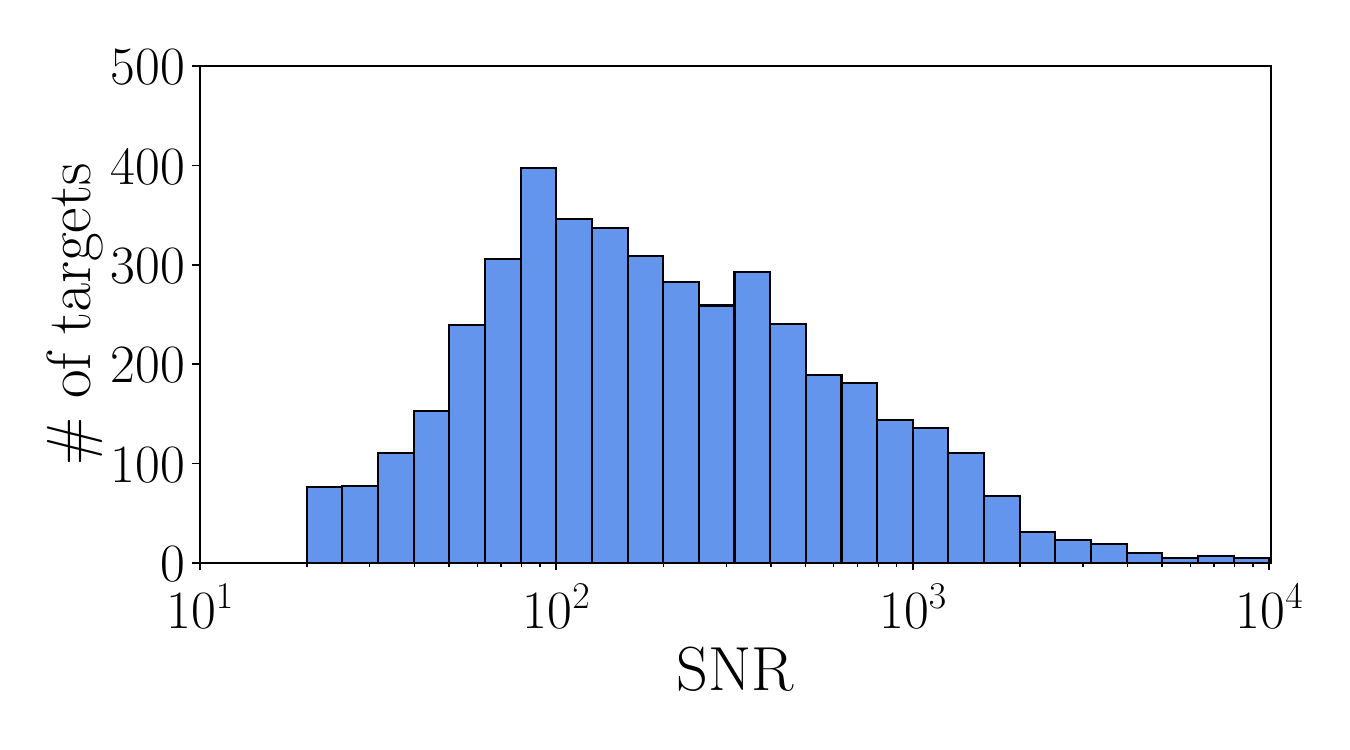}}
  \caption{Distribution of the S/N of the spectral templates used to compute stellar parameters.}
  \label{fig:snrhist}
\end{figure}

The approach developed by \cite{2021A&A...652A.116P} requires precise stellar parameters, namely effective temperature ($T_{\mathrm{eff}}$), surface gravity ($\log{g}$), metallicity ($\mathrm{[Fe/H]}$), and rotational velocity ($v \sin{i}$). In order to ensure that the parameters are derived in as homogeneously a manner as possible, we used the publicly available code Spectroscopic Parameters and atmosphEric ChemIstriEs of Stars {\sc SPECIES}\footnote{\url{https://github.com/msotov/SPECIES}} \citep{2018A&A...615A..76S,2021A&A...647A.157S}. The single-exposure HARPS spectra for each target were shifted to the rest frame and co-added. Only the resulting templates with an S/N~$\geq20$ were then processed with SPECIES, using the default line lists implemented in the code. Fig.~\ref{fig:snrhist} shows the distribution of the S/N of all template spectra.

\begin{figure}
  \resizebox{\hsize}{!}{\includegraphics{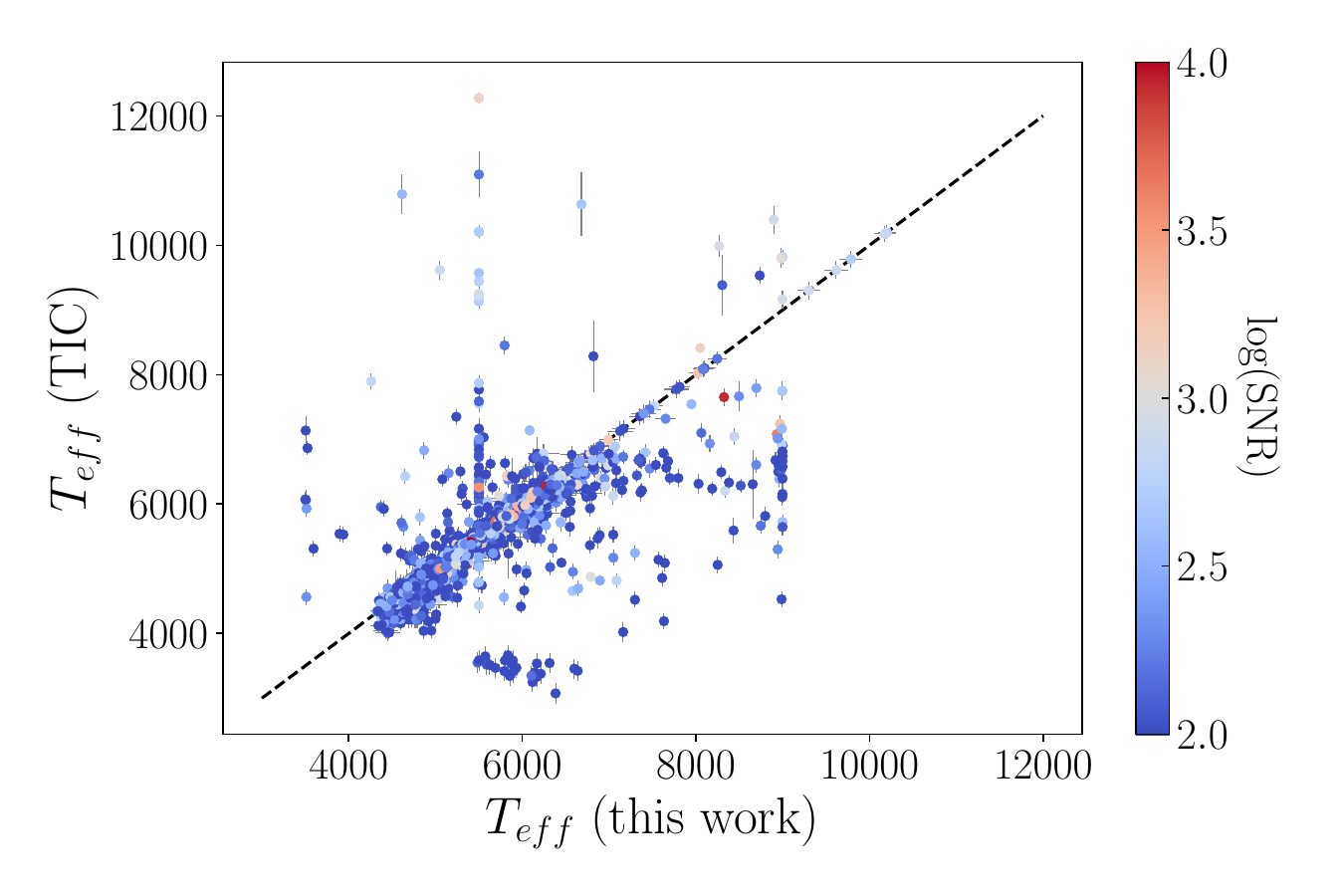}}
  \caption{Comparison of the effective temperature with literature values. The y-axis represents the values published in \cite{2019AJ....158..138S,2021arXiv210804778P}, and the black dashed line represents equality. We note that the apparent tail of deviations towards lower temperatures is comprised of stars for which the RV was not known, and for which the stellar parameter determination failed. All stars with a missing RV value and possible errors in the catalogue are marked via the {\sc vflag} column in our catalogue.}
  \label{fig:teff}
\end{figure}

\begin{figure}
  \resizebox{\hsize}{!}{\includegraphics{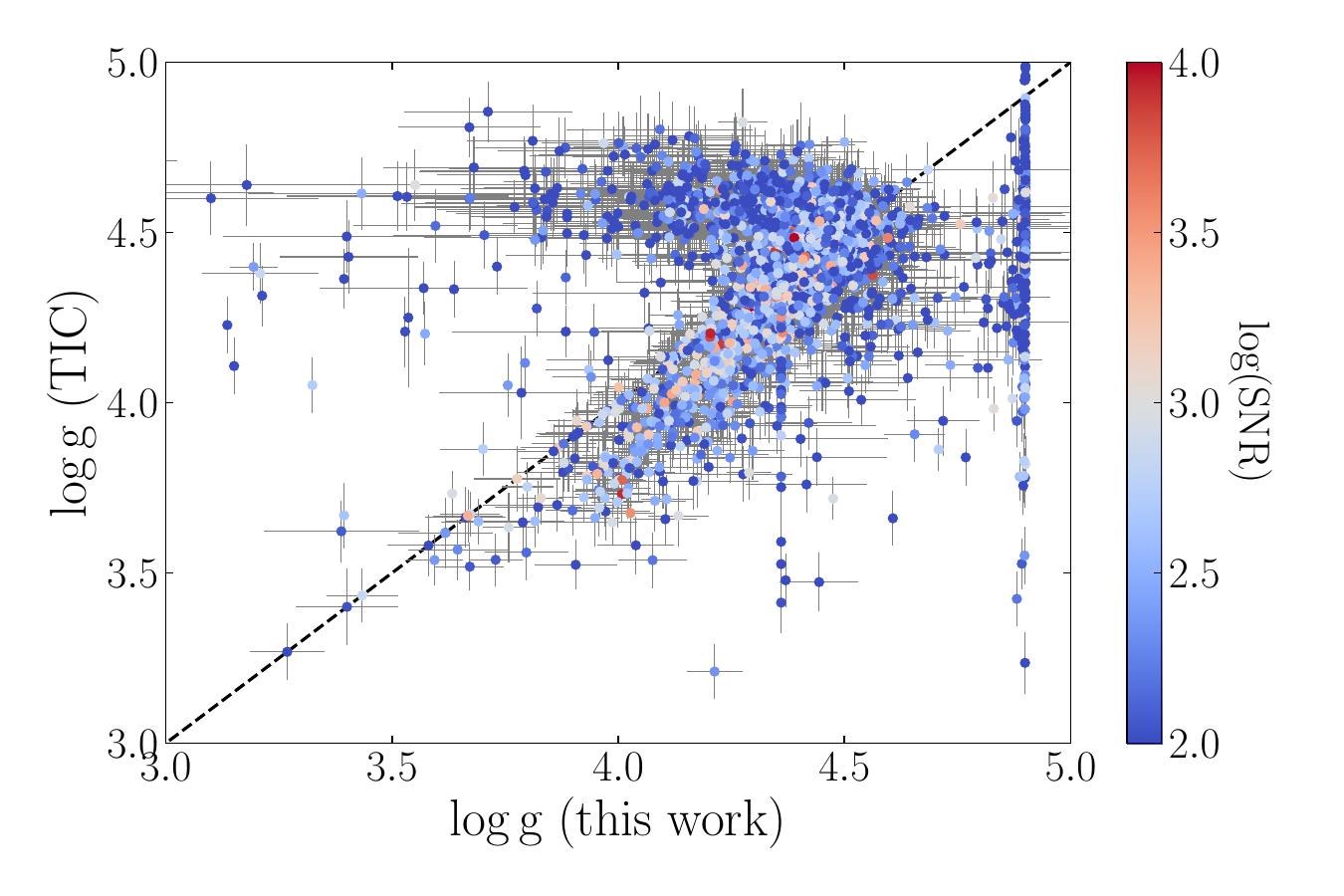}}
  \caption{Comparison of the surface gravity with literature values. The y-axis represents the values published by \cite{2019AJ....158..138S,2021arXiv210804778P}, and the black dashed line represents equality.}
  \label{fig:logg}
\end{figure}

\begin{figure}
  \resizebox{\hsize}{!}{\includegraphics{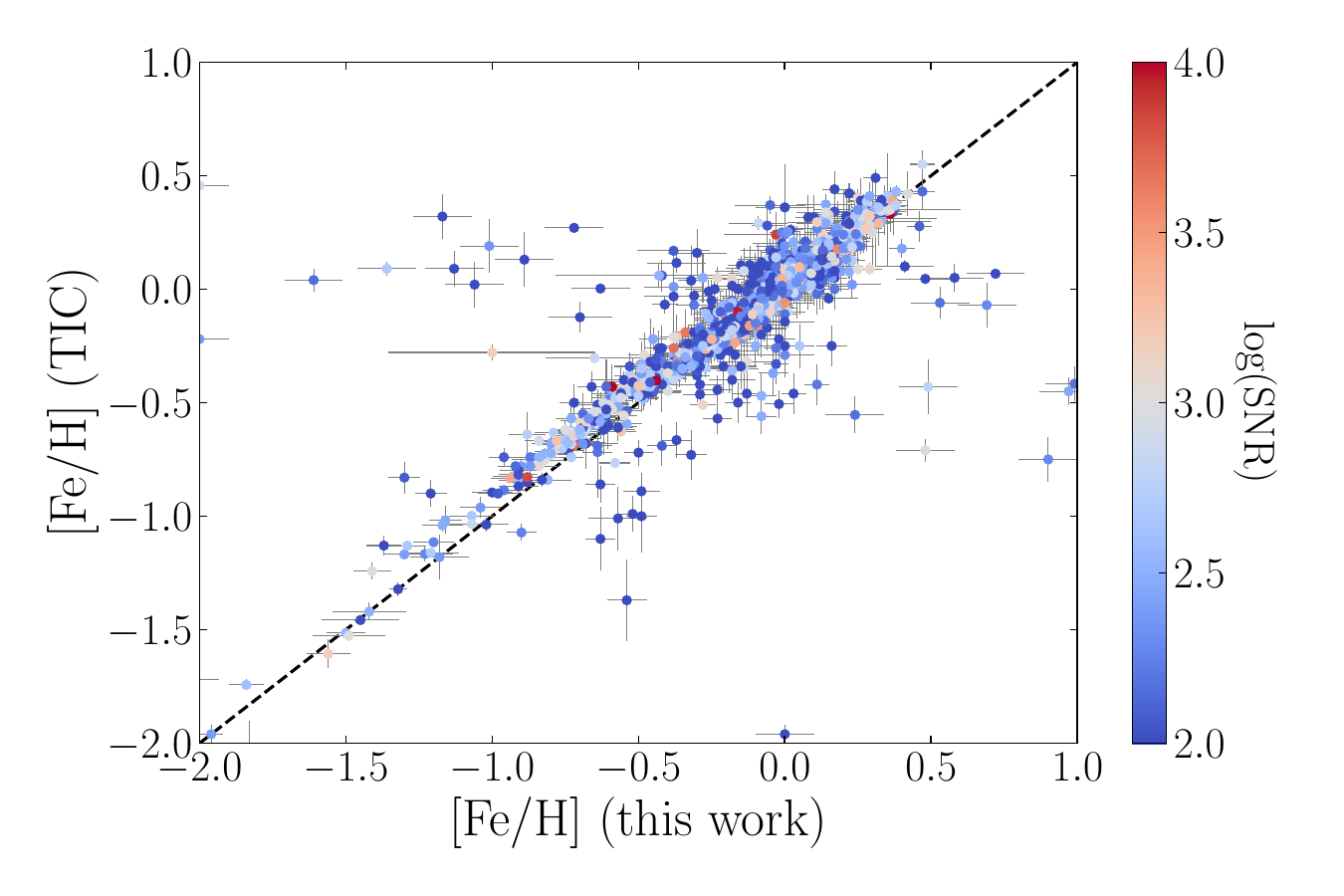}}
  \caption{Comparison of the metallicity with literature values. The y-axis represents the values published by \cite{2019AJ....158..138S,2021arXiv210804778P}, and the black dashed line represents equality.}
  \label{fig:met}
\end{figure}

Since SPECIES is known to converge towards false effective temperatures for stars with $T_{eff}\leq4300$ \citep{2018A&A...615A..76S,2021A&A...647A.157S}, we adopted stellar parameters for all stars with color $B-V>1.24$ from the latest version of the Transiting Exoplanet Survey Satellite (TESS) input catalogue \citep[TIC v8.2][]{2019AJ....158..138S,2021arXiv210804778P}.
Out of a total of 5\,239 stars in the sample, we were able to derive stellar parameters for 3\,230 of them. The discrepancy is caused in most cases by a low S/N of the co-added spectrum, or stellar contamination by a close companion in others. Figures \ref{fig:teff}, \ref{fig:logg}, and \ref{fig:met} show the comparison of the resulting effective temperature, surface gravity, and metallicity with those published in the latest version of the TESS input catalogue.
While there is generally good agreement between the sets of parameters, we do advise that systematic errors in any derivation of stellar parameters may be larger than measured uncertainties \citep{2012ApJ...757..161T}.\newline
The sample of targets for which all stellar parameters have been derived solely based on the HARPS spectra corresponds to a total of 172,948 single spectra, or 68.6\% of the total. In order to increase this fraction, the remaining stars were cross-matched with the TESS input catalogue, which contains the effective temperature, surface gravity, and metallicity, but no rotational velocity. We, therefore, crossmatched the remaining sources with the Gaia DR3 catalogue \citep{2021A&A...649A...1G, 2021A&A...649A...7G}, yielding an additional 54 rotational velocity values. In the stellar parameter catalogue published along with this paper (see Table \ref{table:1}), which is available at the CDS\footnote{\url{https://cdsarc.cds.unistra.fr/viz-bin/cat/J/A+A/vvv/App}}, these targets are marked with $vflag=2$ and those without any rotational velocity with $vflag=1$.

\begin{figure}
  \includegraphics[width=9cm]{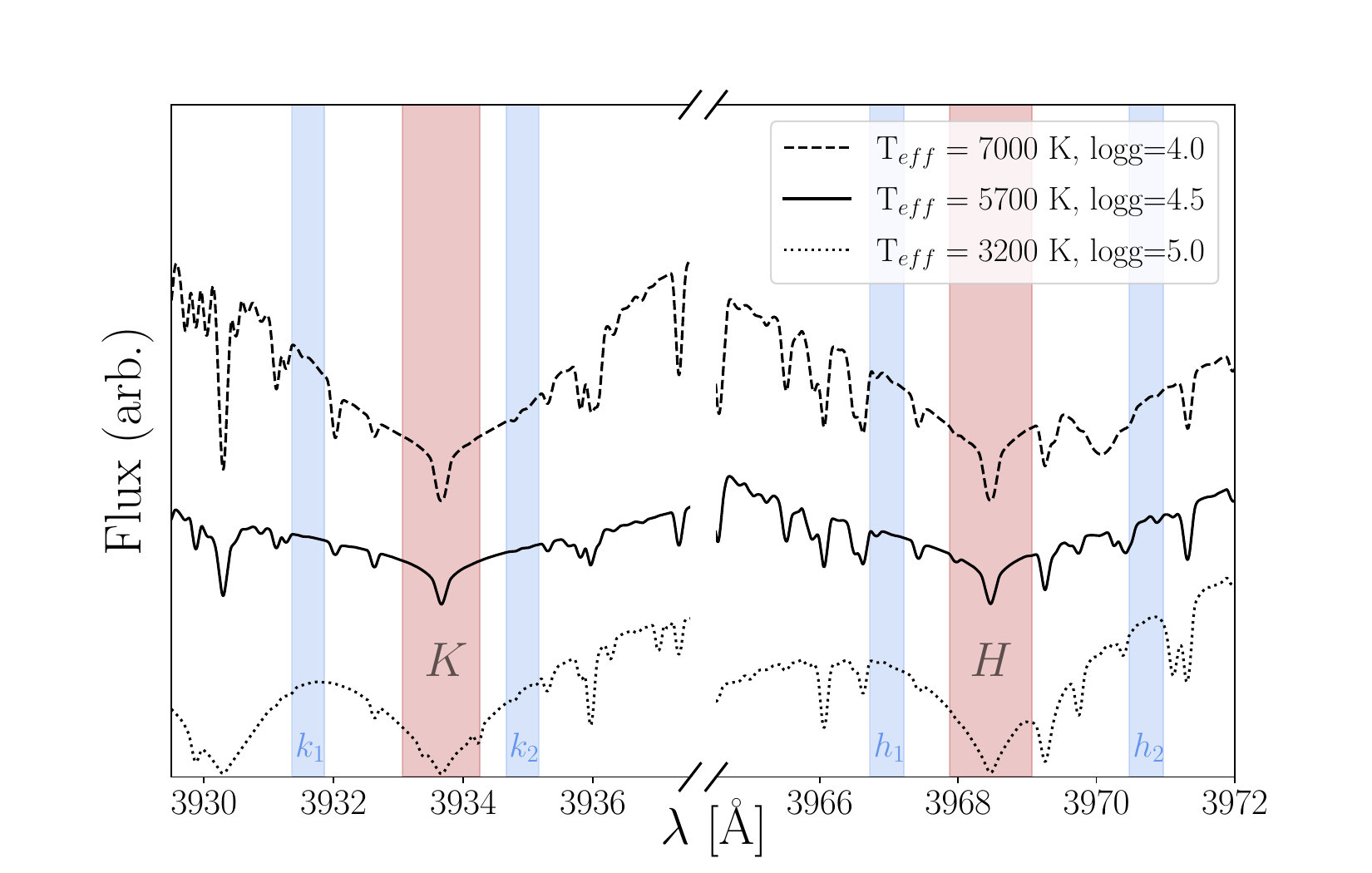}
  \caption{Wavelength bands used for normalization and rectification of the spectra. The black lines show three examples of PHOENIX spectra with different stellar parameters.}
  \label{fig:method}
\end{figure}

\subsection{Extraction of $R_{\mathrm{HK}}^\prime$}
We followed the recipe of \cite{2021A&A...652A.116P}, namely the computation of a grid of fluxes in six bands from PHOENIX models \citep{2013A&A...553A...6H}, with varying $T_{\mathrm{eff}}$, $\log{g}$, and $\mathrm{[Fe/H]}$, where the grid step size follows that of the PHOENIX grid \citep[see Table\,1 in][]{2013A&A...553A...6H}. We adapted the edges of one bandpass ($k_1$) with respect to those used in \cite{2021A&A...652A.116P} in order to minimize the number of absorption lines and hence the influence of errors caused by the determination of stellar metallicity. Tab.~\ref{table:2} lists the updated bandpasses, and fig.~\ref{fig:method} displays the bandpasses for three example model spectra. In addition to the parameters given by the PHOENIX grid, the spectra were artificially broadened with velocities in the range 1-200~km\,s$^{-1}$ in order to account for stellar rotation. The measured spectra were then rectified in the regions of the Ca~{\sc ii} H$\&$K lines and normalized with regard to flux, the photospheric contribution in the lines was subtracted, and the resulting chromospheric excess was normalized with the bolometric flux, yielding $R_{\mathrm{HK}}^\prime$. As mentioned in Sect. \ref{sec:param}, for some stars in the stellar parameter catalogue we were not able to determine $v \sin{i}$. During data reduction, $v \sin{i}$ was set to 2~km\,s$^{-1}$ for these targets. While the time series of these stars can be used to determine rotation periods or activity cycles, $R_{\mathrm{HK}}^\prime$ of these stars may be biased, which is why these entries are marked with the $vflag$ in both the stellar parameter and the time series table.
\begin{table}[t!]
\centering
\small
\caption{Updated wavelength bands used for rectification and flux extraction. The changes with respect to \cite{2021A&A...652A.116P} are highlighted.} 
\label{table:2}
\begin{tabular}{cc}       
\hline\hline
\noalign{\smallskip}
Band&Wavelength range [$\lambda_{\rm min}$, $\lambda_{\rm max}$]$^a$ \\ 
\noalign{\smallskip}
\hline
\noalign{\smallskip}
$K$&[$\lambda_{K}-0.6$~\AA, $\lambda_{K}+0.6$~\AA] \\ 
$H$&[$\lambda_{H}-0.6$~\AA, $\lambda_{H}+0.6$~\AA] \\ 
$\mathbf{k_1}$&[$\lambda_{K}\mathbf{-2.3}$~\AA, $\lambda_{K}\mathbf{-1.8}$~\AA] \\ 
$k_2$&[$\lambda_{K}+2.5$~\AA, $\lambda_{K}+3.0$~\AA] \\ 
$h_1$&[$\lambda_{H}-1.75$~\AA, $\lambda_{H}-1.25$~\AA] \\ 
$h_2$&[$\lambda_{H}+2.0$~\AA, $\lambda_{H}+2.5$~\AA] \\ 
\noalign{\smallskip}
\hline
\end{tabular}
\tablefoot{
\tablefoottext{a}{Central wavelengths: $\lambda_{K}$ = 3933.66\,{\AA} and $\lambda_{H}$~=~3968.47\,{\AA}.}
}
\end{table} 
We determined $R_{\mathrm{HK}}^\prime$ for two sets of spectra: 

{\it (i): Individual measurements.} To provide time series for individual targets, these values and their uncertainties are included as additional columns in the updated {\sc HARPS-RVBank}. Since the purpose of the single-measurement catalogue is the study of the time series of individual stars, $\Delta R_{\mathrm{HK}}^\prime$ was derived purely based on the uncertainty as to the flux.

{\it (ii): Co-added templates.} To obtain an averaged, high-S/N value per target, the resulting $R_{\mathrm{HK}}^\prime$ are listed in the catalogue of stellar parameters described in Sec.~\ref{sec:param}. In order to allow for a comparison of the average activity levels of different stars, here, the uncertainties of the stellar parameters were included in the Monte Carlo approach for the error estimation.

\section{Results and discussion}
\label{sec:3}
Our analysis resulted in a total of 195\,387 $R_{\mathrm{HK}}^\prime$ measurements, or $\sim77\%$ of the entire HARPS {\sc RVBank}. In the following, we give a short discussion of the relationship between $R_{\mathrm{HK}}^\prime$ and RV jitter, and compare our findings to previous results. Table~\ref{table:4} gives an overview of the statistics of the parameter derivation.

\begin{table}[hb!]
\centering
\small
\caption{ Overview of the number of spectra, and the source of the stellar parameters used for the derivation of $R_{\mathrm{HK}}^\prime$.}

\label{table:4}
\begin{tabular}{lc}       
\hline\hline
\noalign{\smallskip}
Sample & number of spectra \\ 
\noalign{\smallskip}
\hline
\noalign{\smallskip}
 HARPS spectra that passed our quality control& 252\,615\\
\noalign{\smallskip}
$R_{\mathrm{HK}}^\prime$ derived with {\sc SPECIES} parameters & 158\,507\\
\noalign{\smallskip}
$R_{\mathrm{HK}}^\prime$ derived with TIC parameters & 36\,880\\
\noalign{\smallskip}
\hline
\noalign{\smallskip}
Total&195\,387\\
\noalign{\smallskip}
\hline
\end{tabular}
\end{table}

\subsection{RV jitter as a function of activity and spectral type}

\begin{figure}
  \resizebox{\hsize}{!}{\includegraphics{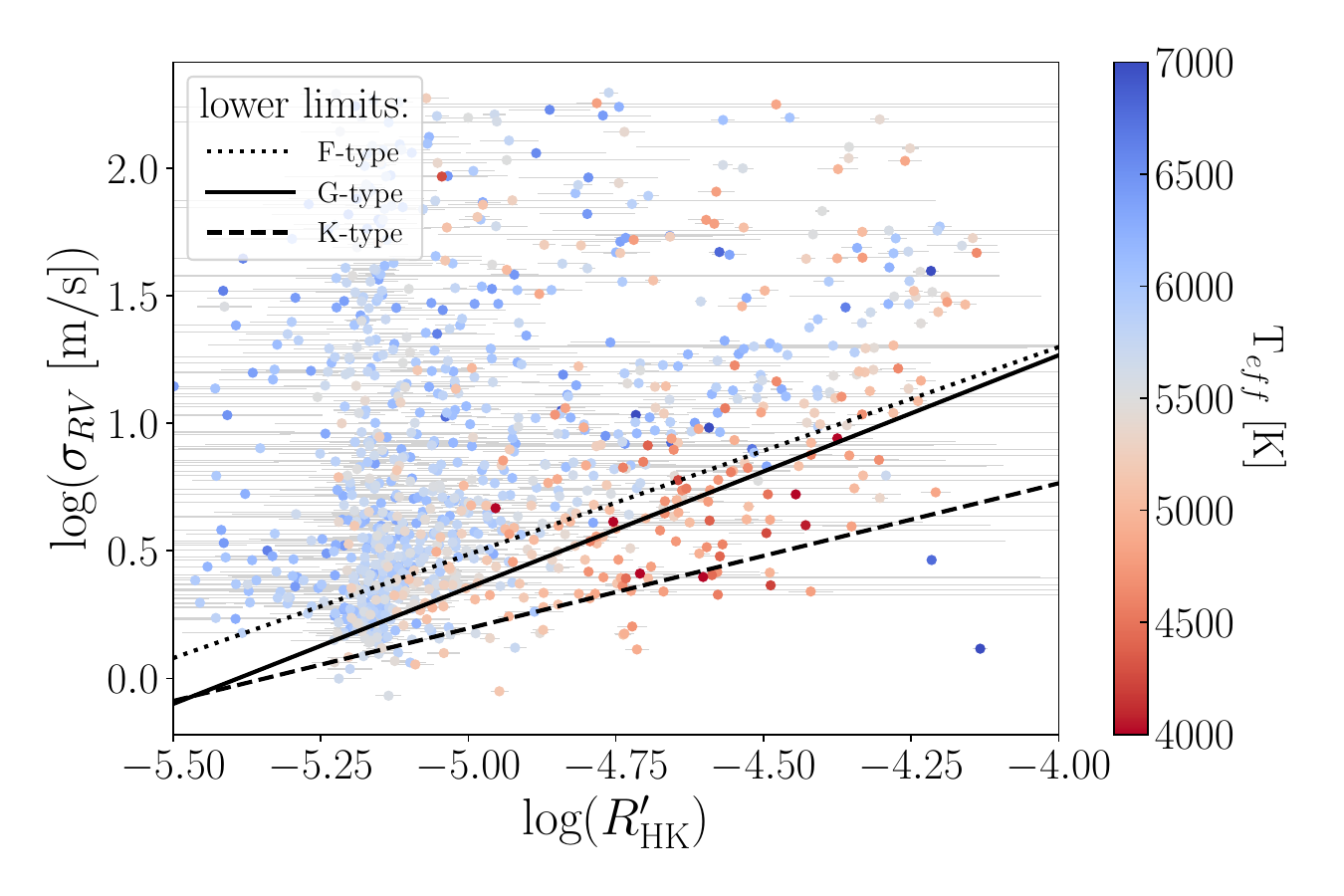}}
  \caption{RV jitter as a function of activity. The colour coding denotes the spectral type, and the dotted, solid, and dashed line correspond to the lower RV jitter for spectral types F, G, and K, respectively.}
  \label{fig:RV_vs_RHKp}
\end{figure}

The number of stars of spectral types F, G, and K in the sample is large enough to derive lower limits on the RV jitter as a function of the spectral type and activity level. Figure~\ref{fig:RV_vs_RHKp} shows the RV jitter $\sigma_{RV}$ as a function of $R_{\mathrm{HK}}^\prime$ derived from the co-added spectra for all main sequence stars of spectral types F, G, or K in the sample with at least 20 individual measurements. Here, we also excluded binaries with an RV jitter of $\sigma_{RV}>200$~m\,s$^{-1}$. For each spectral type, the sample was subdivided into activity levels with step sizes of $0.25$ in log($R_{\mathrm{HK}}^\prime$), and the $10^{th}$ percentile of all RVs was adopted as the lower limit for each bin.

We derived the best-fit linear approximation for the lower limit of RV jitter of F-type stars to be
\begin{equation}
    \log{\sigma_{RV}   \mathrm{[m\,s}^{-1}\mathrm{]}}=0.81\times\log{R_{\mathrm{HK}}^\prime} + 4.55,
\end{equation}
to be
\begin{equation}
    \log{\sigma_{RV}  \mathrm{[m\,s}^{-1}\mathrm{]}}=0.91\times\log{R_{\mathrm{HK}}^\prime} + 4.91
\end{equation}
 for G-type ones, and 
\begin{equation}
    \log{\sigma_{RV}  \mathrm{[m\,s}^{-1}\mathrm{]}}=0.57\times\log{R_{\mathrm{HK}}^\prime} + 3.04
\end{equation}
for K\ dwarfs.

\subsection{Comparison to previous publications}
\begin{figure}
  \resizebox{\hsize}{!}{\includegraphics{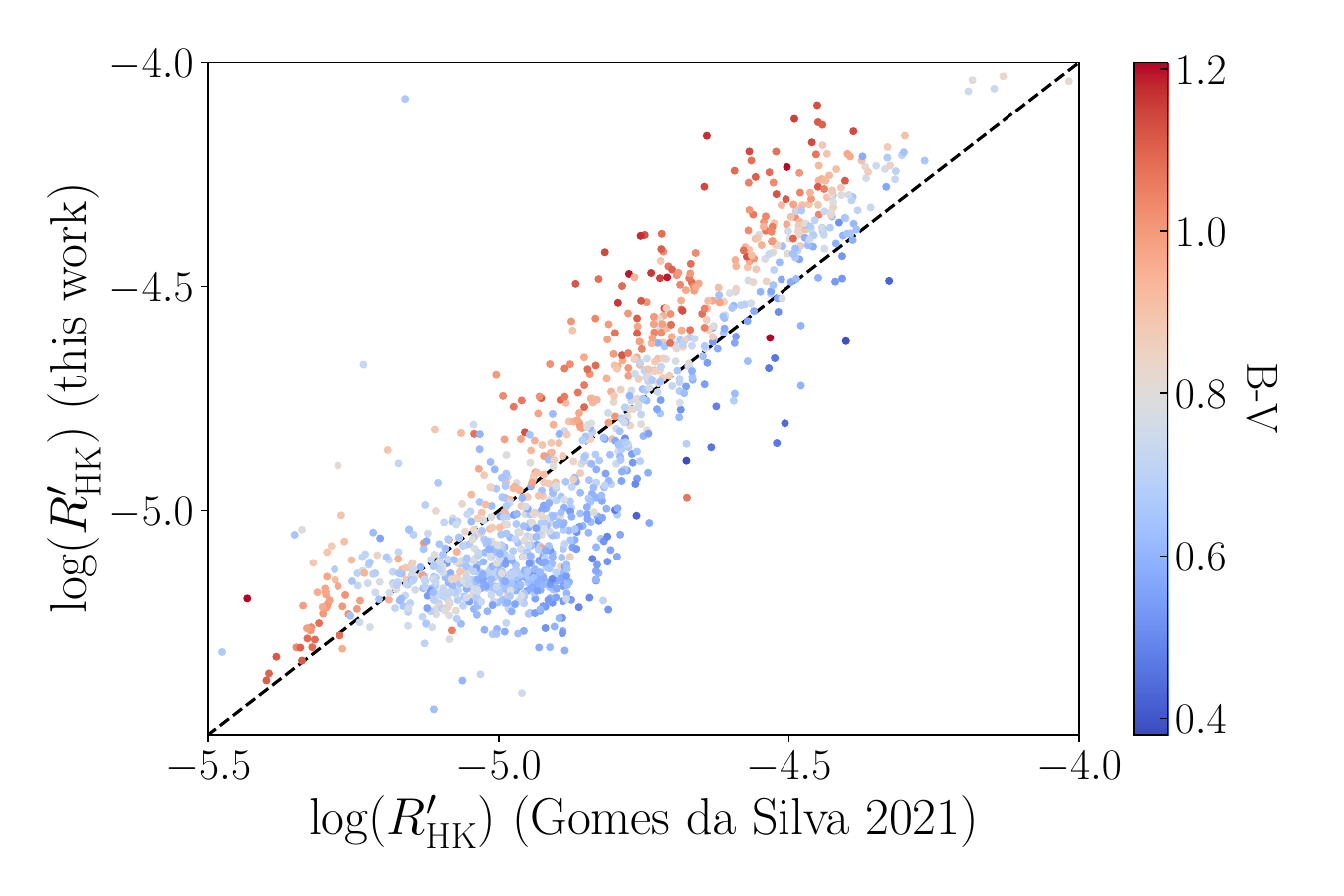}}
  \caption{Comparison of our median $R_{\mathrm{HK}}^\prime$ values to those derived by \cite{2021A&A...646A..77G}. The colour coding denotes B-V.}
  \label{fig:comparison2}
\end{figure}

\cite{2021A&A...646A..77G} recently published a catalogue of chromospheric activity for a sample of 1674 FGK stars based on HARPS data. Since the underlying spectra are entirely included in our dataset, we compared the median $R_{\mathrm{HK}}^\prime$ from their catalogue to that derived from our template spectra. Fig.~\ref{fig:comparison2} shows that there are two types of systematic offset: {\it (i)} Cooler stars appear more active and {\it (ii)} most inactive stars exhibit lower $R_{\mathrm{HK}}^\prime$ in our catalogue, both relative to the values given by \cite{2021A&A...646A..77G}. While identifying the source of these systematics is beyond the scope of this paper, we tentatively conclude that it originates in the different methods for the extraction of $R_{\mathrm{HK}}^\prime$. Our approach takes into account all stellar parameters including metallicity and rotational velocity, whereas the standard method of deriving $R_{\mathrm{HK}}^\prime$, as carried out in \cite{2021A&A...646A..77G}, is the extraction of the Mount Wilson $S$ index, which is subsequently converted to the excess chromospheric emission using a conversion based solely on the stellar colour ($B-V$). 

Neglecting some stellar parameters in extracting $R_{\mathrm{HK}}^\prime$ may be an oversimplification. For example, \citet{2009A&A...493.1099S} found that fast stellar rotation leads to the Ca~{\sc ii} H$\&$K lines to be filled in, thus increasing the flux as determined by the classical Mount Wilson method. Furthermore, it has been found that $R_{\mathrm{HK}}^\prime$ is dependent on the stellar metallicity \citep{2012IAUS..286..335S}. We conclude that any derivation of $R_{\mathrm{HK}}^\prime$ via a conversion factor that depends only on the stellar colour will likely be biased.
Furthermore, as described in \cite{2021A&A...652A.116P}, the method of using narrow bands close to the cores of the Ca~{\sc ii} lines to rectify and normalize the spectra, while resulting in larger statistical errors compared to the classical approach, is less prone to systematics caused by irregularities in the order-merging or normalization of the pipeline-reduced spectra.
\section{Summary and conclusions}
\label{sec:4}
We present an update to the {\sc HARPS-RVBank} with newly calculated RVs, nightly zero points, and an addition of $R_{\mathrm{HK}}^\prime$ for 77\% of all spectra. Table \ref{table:3} shows an example of five (out of $252\,615$) rows of the full catalogue. For 3\,230 stars, we were able to derive stellar parameters using co-added template spectra, which are available as a separate table. The updated RVs will enable the detection of new planet candidates in the same manner as the previous version of the {\sc HARPS-RVBank} \citep[e.g.][]{2020ApJS..250...29F,2021arXiv211209029S}. 
The addition of the relative chromospheric emission in the Ca~{\sc ii}~H\&K lines, $R_{\mathrm{HK}}^\prime$, provides an additional tool for the study of stellar magnetic activity and the identification of false-positive planet detections caused by stellar rotation. The updated {\sc HARPS-RVBank} has already been used in several works on planet detection \citep{2022AJ....164..156T}, stellar activity jitter \citep{2022A&A...666A.143K}, and magnetic activity cycles \citep{2022arXiv221203514F}.
While our $R_{\mathrm{HK}}^\prime$ values show colour-dependent systematic deviations compared to those derived in the classical way, that is, via the S index, we conclude that this is caused by the simplification made in the conversion with a factor that is solely a function of the stellar colour $B-V$.

\renewcommand{\arraystretch}{1.2}
\setlength{\tabcolsep}{10pt}
\begin{table*}[ht!]
\centering
\small
\caption{Updated {\sc HARPS-RVBank}. The dots represent the rest of the columns from the previous version \citep{2020A&A...636A..74T}. The full table is available on the CDS. {The {\sc vflag} and {\sc mflag} columns denote stars for which rotational velocity or metallicity could not be determined. }} 
\label{table:3}
\begin{tabular}{ccccccccc}       
\hline
\hline
Name&RA&Dec&BJD&...&$R_{\mathrm{HK}}^\prime$&$\Delta R_{\mathrm{HK}}^\prime$&$vflag$&$mflag$\\ 
&[$^{\circ}$]&[$^{\circ}$]&[d]&& &\\
\hline
  BD+012494&163.0321195&0.49329&2458249.6510659&...&1.60E-5&1.15E-6&0&0\\
  BD+012494&163.0321195&0.49329&2458251.5921099&...&1.43E-5&9.92E-7&0&0\\
  BD+012494&163.0321195&0.49329&2458262.4807279&...&1.70E-5&1.04E-6&0&0\\
  BD+012494&163.0321195&0.49329&2458263.543513&...&6.60E-6&1.11E-6&0&0\\
\hline
\end{tabular}
\end{table*}

\begin{acknowledgements}
    V.P. acknowledges funding through the Kimmel prize fellowship of the Center for Earth and Planetary Science at the Weizmann Institute of Science, and the Dean's fellowship of the faculty of chemistry (WIS). T.T. acknowledge support by the BNSF programme 'VIHREN-2021' project No. KP-06-DV-5/15.12.2021.
T.T. acknowledges support by the DFG Research Unit FOR 2544 'Blue Planets around Red Stars' project No. KU 3625/2-1. This work was also funded by the Israel Science Foundation through grant No. 1404/22.\\
Based on observations made with ESO Telescopes at the La Silla Paranal Observatory under programme IDs 0100.C-0097, 0100.C-0111, 0100.C-0414, 0100.C-0474, 0100.C-0487, 0100.C-0708, 0100.C-0746, 0100.C-0750, 0100.C-0808, 0100.C-0836, 0100.C-0847, 0100.C-0884, 0100.C-0888, 0100.D-0176, 0100.D-0273, 0100.D-0339, 0100.D-0444, 0100.D-0535, 0100.D-0776, 0101.C-0106, 0101.C-0232, 0101.C-0274, 0101.C-0275, 0101.C-0379, 0101.C-0407, 0101.C-0497, 0101.C-0510, 0101.C-0516, 0101.C-0623, 0101.C-0788, 0101.C-0829, 0101.C-0889, 0101.D-0091, 0101.D-0465, 0101.D-0494, 0101.D-0697, 0102.A-0697, 0102.C-0171, 0102.C-0319, 0102.C-0338, 0102.C-0414, 0102.C-0451, 0102.C-0525, 0102.C-0558, 0102.C-0584, 0102.C-0618, 0102.C-0812, 0102.D-0119, 0102.D-0281, 0102.D-0483, 0102.D-0596, 0103.C-0206, 0103.C-0240, 0103.C-0432, 0103.C-0442, 0103.C-0472, 0103.C-0548, 0103.C-0719, 0103.C-0759, 0103.C-0785, 0103.C-0874, 0103.D-0445, 0104.C-0090, 0104.C-0358, 0104.C-0413, 0104.C-0418, 0104.C-0588, 0104.C-0849, 0104.C-0863, 072.A-0244, 072.C-0096, 072.C-0488, 072.C-0513, 072.C-0636, 072.D-0286, 072.D-0419, 072.D-0707, 073.A-0041, 073.C-0733, 073.C-0784, 073.D-0038, 073.D-0136, 073.D-0527, 073.D-0578, 073.D-0590, 074.C-0012, 074.C-0037, 074.C-0061, 074.C-0102, 074.C-0221, 074.C-0364, 074.D-0131, 074.D-0380, 075.C-0087, 075.C-0140, 075.C-0202, 075.C-0234, 075.C-0332, 075.C-0689, 075.C-0710, 075.C-0756, 075.D-0194, 075.D-0600, 075.D-0614, 075.D-0760, 075.D-0800, 076.C-0010, 076.C-0073, 076.C-0155, 076.C-0279, 076.C-0429, 076.C-0878, 076.D-0103, 076.D-0130, 076.D-0158, 076.D-0207, 077.C-0012, 077.C-0080, 077.C-0101, 077.C-0295, 077.C-0364, 077.C-0513, 077.C-0530, 077.D-0085, 077.D-0498, 077.D-0633, 077.D-0720, 078.C-0037, 078.C-0044, 078.C-0133, 078.C-0209, 078.C-0233, 078.C-0403, 078.C-0510, 078.C-0751, 078.C-0833, 078.D-0067, 078.D-0071, 078.D-0245, 078.D-0299, 078.D-0492, 079.C-0046, 079.C-0127, 079.C-0170, 079.C-0329, 079.C-0463, 079.C-0657, 079.C-0681, 079.C-0828, 079.C-0898, 079.C-0927, 079.D-0009, 079.D-0075, 079.D-0118, 079.D-0160, 079.D-0462, 079.D-0466, 080.C-0032, 080.C-0071, 080.C-0581, 080.C-0664, 080.C-0712, 080.D-0047, 080.D-0086, 080.D-0151, 080.D-0318, 080.D-0347, 080.D-0408, 081.C-0034, 081.C-0119, 081.C-0148, 081.C-0211, 081.C-0388, 081.C-0774, 081.C-0779, 081.C-0802, 081.C-0842, 081.D-0008, 081.D-0065, 081.D-0066, 081.D-0109, 081.D-0531, 081.D-0610, 081.D-0870, 082.B-0610, 082.C-0040, 082.C-0212, 082.C-0308, 082.C-0312, 082.C-0315, 082.C-0333, 082.C-0357, 082.C-0390, 082.C-0412, 082.C-0427, 082.C-0608, 082.C-0718, 082.D-0499, 082.D-0833, 083.C-0186, 083.C-0413, 083.C-0627, 083.C-0794, 083.C-1001, 083.D-0040, 083.D-0549, 083.D-0668, 083.D-1000, 084.C-0185, 084.C-0228, 084.C-0229, 084.C-1024, 084.C-1039, 084.D-0338, 084.D-0591, 085.C-0019, 085.C-0063, 085.C-0318, 085.C-0393, 085.C-0614, 085.D-0296, 085.D-0395, 086.C-0145, 086.C-0230, 086.C-0284, 086.C-0448, 086.D-0078, 086.D-0240, 086.D-0657, 087.C-0012, 087.C-0368, 087.C-0412, 087.C-0497, 087.C-0649, 087.C-0831, 087.C-0990, 087.D-0511, 087.D-0771, 087.D-0800, 088.C-0011, 088.C-0323, 088.C-0353, 088.C-0513, 088.C-0662, 088.D-0066, 089.C-0006, 089.C-0050, 089.C-0151, 089.C-0415, 089.C-0497, 089.C-0732, 089.C-0796, 089.D-0138, 089.D-0302, 089.D-0383, 090.C-0131, 090.C-0395, 090.C-0421, 090.C-0540, 090.C-0849, 090.D-0256, 091.C-0034, 091.C-0184, 091.C-0271, 091.C-0438, 091.C-0456, 091.C-0471, 091.C-0844, 091.C-0853, 091.C-0866, 091.C-0936, 091.D-0469, 091.D-0759, 091.D-0836, 092.C-0282, 092.C-0427, 092.C-0454, 092.C-0579, 092.C-0715, 092.C-0721, 092.C-0832, 092.D-0206, 092.D-0261, 092.D-0363, 093.C-0062, 093.C-0163, 093.C-0184, 093.C-0376, 093.C-0409, 093.C-0417, 093.C-0423, 093.C-0474, 093.C-0540, 093.C-0919, 093.D-0367, 093.D-0833, 094.C-0090, 094.C-0297, 094.C-0322, 094.C-0428, 094.C-0790, 094.C-0797, 094.C-0894, 094.C-0901, 094.C-0946, 094.D-0056, 094.D-0274, 094.D-0596, 094.D-0704, 095.C-0040, 095.C-0105, 095.C-0367, 095.C-0551, 095.C-0718, 095.C-0799, 095.C-0947, 095.D-0026, 095.D-0155, 095.D-0194, 095.D-0269, 095.D-0717, 096.C-0053, 096.C-0082, 096.C-0183, 096.C-0210, 096.C-0331, 096.C-0417, 096.C-0460, 096.C-0499, 096.C-0657, 096.C-0708, 096.C-0762, 096.C-0876, 096.D-0064, 096.D-0072, 096.D-0257, 096.D-0402, 097.C-0021, 097.C-0090, 097.C-0277, 097.C-0390, 097.C-0434, 097.C-0561, 097.C-0571, 097.C-0864, 097.C-0948, 097.C-1025, 097.D-0120, 097.D-0150, 097.D-0156, 097.D-0420, 098.C-0042, 098.C-0269, 098.C-0292, 098.C-0304, 098.C-0366, 098.C-0440, 098.C-0446, 098.C-0518, 098.C-0645, 098.C-0739, 098.C-0820, 098.C-0860, 098.D-0187, 099.C-0081, 099.C-0093, 099.C-0138, 099.C-0205, 099.C-0303, 099.C-0304, 099.C-0334, 099.C-0374, 099.C-0458, 099.C-0491, 099.C-0599, 099.C-0798, 099.C-0880, 099.C-0898, 099.D-0236, 099.D-0380, 105.2045.001, 105.2045.002, 105.207T.001, 105.208G.001, 105.20AK.002, 105.20AZ.001, 105.20B1.003, 105.20FX.001, 105.20G9.001, 105.20GX.001, 105.20L0.001, 105.20L8.002, 105.20MP.001, 105.20N0.001, 105.20NV.001, 105.20PH.001, 106.20Z1.001, 106.20Z1.002, 106.212H.001, 106.212H.007, 106.212H.009, 106.212H.010, 106.215E.001, 106.215E.002, 106.215E.004, 106.216H.001, 106.21DB.001, 106.21DH.001, 106.21ER.001, 106.21GB.003, 106.21MA.001, 106.21PJ.001, 106.21PJ.002, 106.21R4.001, 106.21TJ.001, 106.21TJ.008, 107.22R1.001, 107.22UN.001, 108.21XB.001, 108.21YY.001, 108.21YY.002, 108.21YY.004, 108.21YY.005, 108.222V.001, 108.2271.001, 108.2271.002, 108.2271.003, 108.229Z.001, 108.229Z.002, 108.22A8.001, 108.22CE.001, 108.22E7.001, 108.22KV.001, 108.22KV.002, 108.22KV.003, 108.22KV.004, 108.22KV.005, 108.22L8.001, 108.22LE.001, 108.22LR.001, 108.23MM.004, 108.23MM.005, 109.230J.001, 109.2317.001, 109.233Q.001, 109.2374.001, 109.2392.001, 109.239V.001, 109.23J8.001, 110.23XW.001, 110.23XW.002, 110.23YQ.001, 110.241K.001, 110.241K.002, 110.242T.001, 110.2434.001, 110.2438.001, 110.2460.001, 110.248C.001, 110.24BB.001, 110.24C8.001, 110.24C8.002, 110.24D8.001, 1101.C-0557, 1101.C-0721, 1102.C-0249, 1102.C-0339, 1102.C-0923, 1102.D-0954, 111.24PJ.002, 111.24UR.001, 111.24ZQ.001, 111.24ZW.001, 111.250B.001, 111.254A.001, 111.254E.001, 111.254R.002, 111.255C.001, 111.255C.003, 111.255C.004, 111.263V.001, 111.264X.001, 180.C-0886, 182.D-0356, 183.C-0437, 183.C-0972, 183.D-0729, 184.C-0639, 184.C-0815, 185.D-0056, 187.D-0917, 188.C-0265, 188.C-0779, 190.C-0027, 190.D-0237, 191.C-0505, 191.C-0873, 191.D-0255, 192.C-0224, 192.C-0852, 196.C-0042, 196.C-1006, 198.C-0169, 198.C-0836, 198.C-0838, 2101.C-5015, 276.C-5009, 281.D-5052, 281.D-5053, 281.D-5054, 282.C-5034, 282.C-5036, 282.D-5006, 283.C-5017, 283.C-5022, 288.C-5010, 289.C-5053, 289.D-5015, 292.C-5004, 295.C-5031, 295.C-5035, 297.C-5051, 495.L-0963, 60.A-9036, 60.A-9109, 60.A-9501, 60.A-9700, 60.A-9709.\\
This work has made use of data from the European Space Agency (ESA) mission
{\it Gaia} (\url{https://www.cosmos.esa.int/gaia}), processed by the {\it Gaia}
Data Processing and Analysis Consortium (DPAC,
\url{https://www.cosmos.esa.int/web/gaia/dpac/consortium}). Funding for the DPAC
has been provided by national institutions, in particular the institutions
participating in the {\it Gaia} Multilateral Agreement.
\end{acknowledgements}

\bibliographystyle{aa}
\bibliography{astro}


\end{document}